\documentclass[twocolumn, prx, reprint, amsfonts, amsmath, amssymb, superscriptaddress, longbibliography, floatfix,aps]{revtex4-1}
\usepackage[T1]{fontenc}
\usepackage{graphicx}
\usepackage[dvipsnames]{xcolor}
\usepackage[colorlinks,linkcolor=Blue,citecolor=Blue]{hyperref}
\usepackage{braket}
\usepackage{enumitem}
\usepackage[english]{babel}
\usepackage{titlesec}

\usepackage{pifont}
\usepackage{fontawesome} 
\usepackage{MnSymbol}
\usepackage{array, makecell}
\newcolumntype{L}[1]{>{\raggedright\let\newline\\\arraybackslash\hspace{0pt}}m{#1}}
\newcolumntype{C}[1]{>{\centering\let\newline\\\arraybackslash\hspace{0pt}}m{#1}}
\newcolumntype{R}[1]{>{\raggedleft\let\newline\\\arraybackslash\hspace{0pt}}m{#1}}

\begin{document}
\title{Superconducting Photocurrent and Light Enriched Supercurrent Phase Relation}
\author{Oles Matsyshyn}
\email{oles.matsyshyn@ntu.edu.sg}
\affiliation{Division of Physics and Applied Physics, School of Physical and Mathematical Sciences, Nanyang Technological University, Singapore 637371}
\author{Justin C. W. Song}
\email{justinsong@ntu.edu.sg}
\affiliation{Division of Physics and Applied Physics, School of Physical and Mathematical Sciences, Nanyang Technological University, Singapore 637371}
\date{\today}

\begin{abstract}
Noncentrosymmetric superconductors are expected to exhibit DC photocurrents even for irradiation frequencies below the superconducting gap. Such superconducting photocurrent are non-dissipative and track the quantum geometry of the superconducting state. Here we argue that superconducting photocurrent drives changes to the constitutive superconducting current phase relation (CPR) manifesting in a light controlled inductive response as well as altering the critical current. For chiral incident light, we find that superconducting photocurrent can transform reciprocal CPR into a non-reciprocal CPR producing a non-reciprocal inductance and light-controlled superconducting diode effect. These provide a protocol for measuring superconducting photocurrent and new tools for mapping the quantum geometry and order parameter of superconductors.
\end{abstract}

\maketitle

Photocurrent is an essential characteristic of opto-electronic devices \cite{SzeBook,NelsonBook}. Commonly found at extrinsic p-n junction interface where its presence changes the current-voltage characteristics of the device \cite{SzeBook,NelsonBook}, DC photocurrent can even persist without a p-n junction in noncentrosymmetric materials \cite{von1981theory,sipe2000second,morimotoSciAdv, PhysRevX.10.041041, Ma2023}. In these noncentrosymmetric materials, photocurrent instead depends on the quantum geometry of bandstructure that include the shift vector in semiconductors \cite{von1981theory,sipe2000second,morimotoSciAdv}, or the Berry curvature dipole in metallic systems \cite{PhysRevLett.105.026805,SodemannFu, matsyshyn2019nonlinear,parker2019diagrammatic}. 

A particularly interesting example is that of photocurrent in noncentrosymmetric superconductors~\cite{PhysRevB.105.024308,Tanaka_2023,matsyshynSBCD}. In these, light irradiation underneath the superconducting gap can produce a non-dissipative DC photocurrent without breaking Cooper pairs or degrading the superconducting order~\cite{PhysRevB.105.024308,Tanaka_2023,matsyshynSBCD}. Even as such superconducting photocurrent readily flow in noncentrosymmetric superconductors, measuring its value is challenging. For example, because a superconductor cannot sustain a DC voltage drop, when it is connected to an external circuit (formed out of a normal metal) for measurement, DC superconducting photocurrents are readily screened, perfectly compensated to ensure a zero voltage drop across the superconductor. How do we extract the superconducting photocurrent in a superconducting device?

\begin{figure}
\centering
\includegraphics[width=0.49\textwidth]{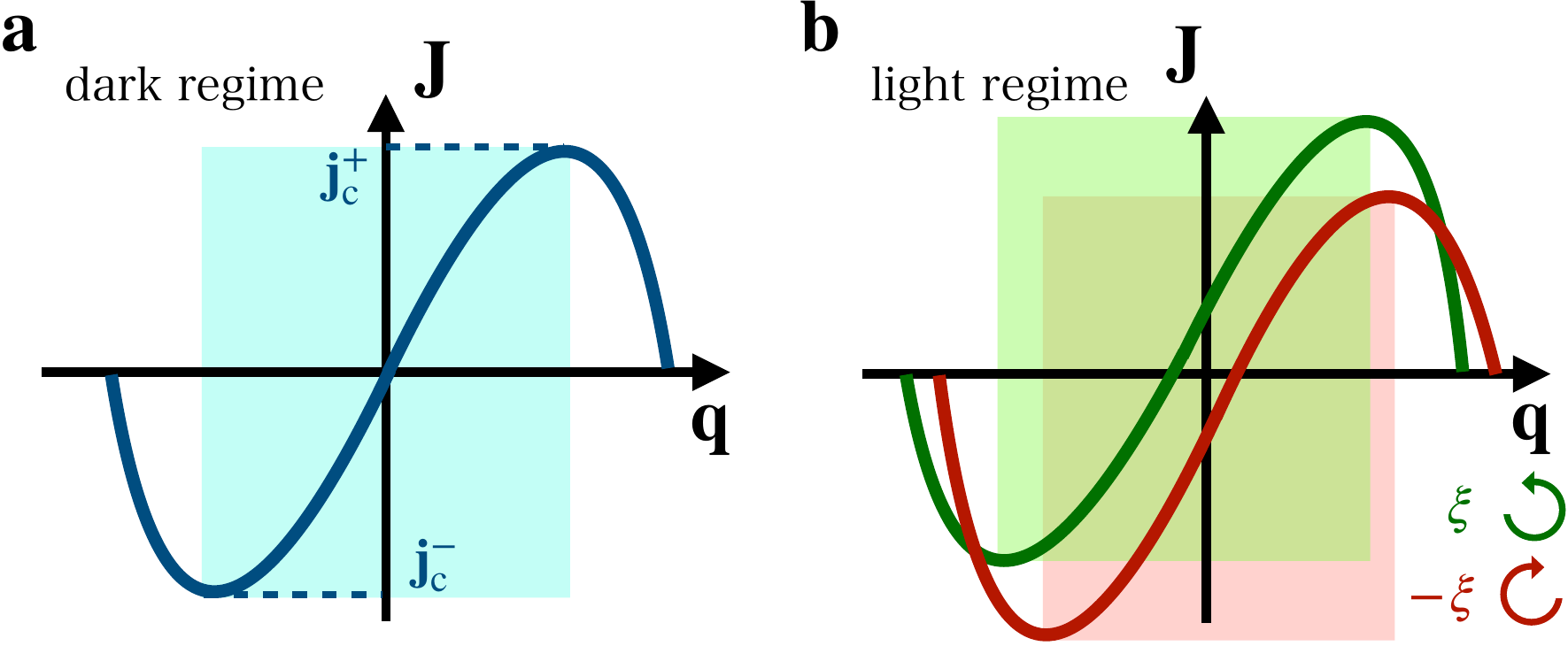}
\caption{\textbf{Bulk superconducting current phase relation in a non-magnetic superconductor.} {\bf a}. In the absence light-induced photocurrent (dark regime), the superconducting current phase relation (CPR) is odd in the center of mass momentum of the Cooper pairs $\hbar q$. This produces a reciprocal superfluid weight $\rho (\mathbf{q}) = \rho(-\mathbf{q})$ and reciprocal critical current $\mathbf{j}_c^+ = - \mathbf{j}_c^-$. {\bf b.} Irradiation by light can enrich CPR (light regime): in the presence of circularly polarized light induced photocurrennts, the CPR of non-centrosymmetric superconductors can be shifted and acquire a non-reciprocity. This produces a non-reciprocal superfluid weight $\rho (\mathbf{q}) \neq \rho(-\mathbf{q})$ and imbalances the critical current $[\mathbf{j}_c^+]_{\rm cir} \neq - [\mathbf{j}_c^-]_{\rm cir}$.} 
\label{SDEFIG1}
\end{figure}

Here we argue that superconducting photocurrent drives changes to the superconducting current phase relation (CPR) in the superconductor Fig.~\ref{SDEFIG1}. In the absence of light irradiation, dark CPR tracks how the center of mass momentum of the Cooper pairs (phase gradient) produces supercurrent characterizing its inductive response \cite{tinkham2004introduction} as well as critical current \cite{Daido_2022}. When a superconducting photocurrent develops, we find that light CPR shifts producing a distinct light-enriched CPR, Fig.~\ref{SDEFIG1}b.

Importantly, as we will see below, light-enriched CPR transforms its inductive response and alters the critical current. We find both light induced changes to inductance and critical currents are driven by superconducting photocurrent and reach appreciable values in a range of noncentrosymmetric superconducting devices. Indeed, we expect that such light-enriched CPR can be readily realized in a range of noncentrosymmetric superconducting devices, e.g., superconducting rhombohedral trilayer graphene \cite{Zhou2021}. Together these provide a protocol to detect the superconducting photocurrent and its associated quantum geometry. Interesting, such light-induced non-reciprocity may be used for dissipationless nonlinearities in superconducting devices~\cite{Fatemi.21.064029} and join a host of light-induced protocols in controlling the behavior of superconductors \cite{Mitrano2016,sciadv.aau6969,PhysRevLett.122.133602,Luo2023,arora2025chiralcavitycontrolsuperconducting}. 

\vspace{2mm}
\textit{\color{blue}{Current phase relationship and photocurrent.}} We begin by examining a superconductor in the presence of an oscillating electromagnetic field $\mathbf{A}(t)$ via a mean-field Bogoliubov de Gennes (BdG) Hamiltonian, \( \hat{H}\hspace{-0.75mm} =\hspace{-0.75mm} \frac{1}{2} \sum_{\mathbf{k}} \Psi^\dagger_{\mathbf{k,q}} \hat{H}^{\rm BdG}_{\mathbf{k,q}}[\mathbf{A}(t)] \Psi_{\mathbf{k,q}},\) in the Bloch-Nambu basis with  
\begin{equation}
\hat{H}^{\rm BdG}_{\mathbf{k,q}}[\mathbf{A}(t)] = 
\begin{pmatrix}
\hat{H}^{(0)}_{\mathbf{k+q}}[\mathbf{A}(t)] & \hat{\Delta} \\
\hat{\Delta}^\dagger & -\big\{\hat{H}^{(0)}_{\mathbf{-k+q}}[\mathbf{A}(t)]\big\}^T
\end{pmatrix},
\label{eq:BdGH}
\end{equation}
where the Bloch Hamiltonian of the parent metal is \( \hat{H}_{\mathbf{k}}^{(0)}[0] = \hat H_0(\mathbf{k}) - \mu \mathbb{I} \) with $\mathbf{k}$ the Bloch momentum, \( \hat{\Delta} \) the local pairing operator, and \( \mathbf{q} \) is the Cooper pair momentum. Here \( \mu \) is the chemical potential and \( \hat H_{\mathbf{k}}^{(0)}[0] \ket{n, \mathbf{k}} = \epsilon_{\mathbf{k}} \ket{n, \mathbf{k}} \) with $\epsilon_{\mathbf{k}}$ and $\ket{n, \mathbf{k}}$ the energy and eigenstates of the parent metal tracking single electron quasiparticles. In contrast, the eigenstates \( \ket{\Psi_{\mathbf{k,q}}} \) of the BdG Hamiltonian describe Bogoliubov quasiparticles in the superconducting state: $\hat{H}^{\rm BdG}_{\mathbf{k,q}}[0]\ket{\Psi^a_{\mathbf{k,q}}} = \varepsilon_{\mathbf{k},a} (\mathbf{q})\ket{\Psi^a_{\mathbf{k,q}}}$. 

At the core of supercurrent dynamics is the current phase relation (CPR) that links the supercurrent to the spatial phase gradient sustained across a device. In a bulk superconductor, such a phase gradient is directly proportional to the collective center of mass momentum of the Cooper pairs $2\hbar \mathbf{q}$, and can be expressed as:
\begin{equation}\label{fullJ}
    \mathbf{j}[\mathbf{q},t] \hspace{-0.5mm}=\hspace{-0.5mm}\int_{\mathbf{k}}{\rm Tr}\{\hat{\rho}_{\mathbf{k,q}}[\mathbf{A}(t)]~e\hat{\mathbf{v}}_{\mathbf{k,q}}[\mathbf{A}(t)]\},
\end{equation}
where $e\hat{\mathbf{v}}_{\mathbf{k,q}}\hspace{-1mm} =\hspace{-1mm} [\partial_{\mathbf{A}}\hspace{-0.5mm}\hat{H}^{(0)}_{\mathbf{k+q}}[\hspace{-0.5mm}\mathbf{A}\hspace{-0.5mm}(t)],\hat{0};\hat{0},-\partial_{\mathbf{A}}\hspace{-0.5mm}\big\{\hspace{-0.5mm}\hat{H}^{(0)}_{\mathbf{-k+q}}[\hspace{-0.5mm}\mathbf{A}\hspace{-0.5mm}(t)]\big\}\hspace{-0.5mm}^T\hspace{-0.5mm}]$ is the current operator and $\hat{\rho}_{\mathbf{k,q}}[\mathbf{A}(t)]$ is the density matrix of the Bogoliubov quasiparticles. 

CPR takes on a familiar form in the \textit{dark} regime (i.e in absence of the electromagnetic fields $\mathbf{A} (t) = 0$). Evaluating $\bra{\Psi^a_{\mathbf{k,q}}} \hat{\rho}_{\mathbf{k,q}}[0]\ket{\Psi^b_{\mathbf{k,q}}} = f[\varepsilon_{\mathbf{k},a}(\mathbf{q})]\delta_{ab}\equiv f_a\delta_{ab}$ where $f(x) = [1+\exp(\beta x)]^{-1}$ is the Fermi function with $\beta = (k_B T)^{-1}$,  Eq.(\ref{fullJ}) naturally reduces to 
\begin{equation}\label{jdark}
    \mathbf{j}_{\rm dark}[\mathbf{q}] =  \frac{e}{\hbar}\int_{\mathbf{k}}\sum_{a}f_a\partial_\mathbf{q} \varepsilon_{\mathbf{k},a}(\mathbf{q}), 
\end{equation}
where we have used $\partial_\mathbf{A}\equiv \frac{e}{\hbar}\partial_{\mathbf{q}}$~\cite{matsyshynSBCD}. For small $\mathbf{q}$, Eq.~(\ref{jdark}) can be linearized as $ \delta {j}^\alpha [\mathbf{q}] =\frac{e}{\hbar}\rho^{\alpha \beta}_{\rm dark }(q^\beta-q^\beta_0)$ via the dark superfluid weight $\rho^{\alpha \beta}_{\rm dark } = \int_{\mathbf{k}} \sum_{a}f_a\partial_{\mathbf{q}_\alpha} \partial_{\mathbf{q}_\beta} \varepsilon_{\mathbf{k},a}(\mathbf{q}) $ that tracks the slope of the dark CPR. This illustrates how momentum of the Cooper pairs directly produces a supercurrent.

Symmetry plays a critical role in the behavior of the dark CPR in Eq.~(\ref{jdark}). For e.g., in the presence of either inversion (IS) or time-reversal (TRS) symmetries: \( \varepsilon_{\mathbf{k},a}(\mathbf{q}) = \varepsilon_{-\mathbf{k},a}(-\mathbf{q}) \), resulting in an odd parity $\mathbf{j}_{\rm dark}[\mathbf{q}] = -\mathbf{j}_{\rm dark}[-\mathbf{q}]$ as shown in Fig.~\ref{SDEFIG1}a; similarly, the superfluid weight is $\mathbf{q}$-even function $\rho_{\rm dark }[\mathbf{q}]=\rho_{\rm dark }[-\mathbf{q}]$ in Fig.~\ref{SDEFIG1}b. These mean that the supercurrent dynamics are reciprocal for both forward and backward propagation with the same magnitude of critical current, Fig.~\ref{SDEFIG1}a.

As we now argue, oscillating electromagnetic fields $\mathbf{E}_{\rm drive} (t)=-\partial_t\mathbf{A}(t)=\boldsymbol{\mathcal{E}}e^{-i\omega_{d} t}+c.c.,$ can dramatically change the CPR even in the dissipationless limit, where frequencies $\hbar \omega_d$ are smaller that the superconducting gap $\Delta$. In this limit, the dynamics of the current can be analyzed within an adiabatic approach \cite{Liang_2017}. Focusing on the DC component of the photoinduced supercurrent in Eq.~(\ref{fullJ}) by averaging across one cycle we have \cite{SodemannFu,matsyshyn2019nonlinear,shi2023berry}
\begin{equation}\label{jphoto}
     {j}^\gamma_{\rm photo} [\mathbf{q}] \hspace{-0.75mm}= \hspace{-0.75mm}\frac{2e^3}{\hbar }\bigg\{\hspace{-0.5mm}\frac{\mathcal{E}^{*\alpha} \mathcal{E}^\beta}{\hbar^2\omega^2_{d}}{\mathcal{ J}}^{\alpha\beta\gamma}_{\mathbf{q}}  \hspace{-0.5mm}+\hspace{-0.5mm}\frac{{\rm Im}[\mathcal{E}^{*\alpha} \mathcal{E}^\beta]}{\hbar\omega_{d}}  \mathcal{D}^{\alpha\beta\gamma}_{\mathbf{q}}\hspace{-0.5mm}\bigg\},
\end{equation}
are DC dissipationless photocurrents produced in the superconductor. \(\mathcal{ J}^{\alpha\beta\gamma}_{\mathbf{q}} \hspace{-1mm}=\hspace{-1mm}\int_{\mathbf{k}}\sum_{a}f_a\partial^\gamma_{\mathbf{q}}\partial^\beta_{\mathbf{q}} \partial^\alpha_{\mathbf{q}}\varepsilon_{\mathbf{k},a} (\mathbf{q})\) is a superconducting Jerk contribution \cite{Kitamura_2023,PhysRevB.105.024308,Tanaka_2023,matsyshynSBCD} and $\mathcal{D}^{\alpha\beta\gamma}_{\mathbf{q}}\hspace{-1mm}= \int_{\mathbf{k}}\sum_{a}f_a\partial^\alpha_\mathbf{q}[\partial_{\mathbf{q}}^\beta {\mathcal{A}}^\gamma_{\mathbf{k},aa}(\mathbf{q})\hspace{-1mm}-\hspace{-1mm}\partial_{\mathbf{q}}^\gamma {\mathcal{A}}^\beta_{\mathbf{k},aa}(\mathbf{q})]$ is a superconducting Berry curvature dipole \cite{Kitamura_2023,PhysRevB.105.024308,Tanaka_2023,matsyshynSBCD}. Here $\boldsymbol{\mathcal{A}}_{\mathbf{k},aa}(\mathbf{q})=\bra{{\Psi}^a_{\mathbf{k,q}}}i\boldsymbol{\partial}_{\mathbf{q}}{\Psi}^a_{\mathbf{k,q}}\rangle$ is a Bogoliubov Berry connection. While Jerk contributions are produced by linearly polarized light since $\mathcal{ J}^{\alpha\beta\gamma}_{\mathbf{q}}$ is symmetric under permutation of indices, superconducting Berry curvature dipole induced contributions are produced by circularly polarized light. 

As second-order nonlinearities, both contributions to Eq.~(\ref{jphoto}) require broken inversion symmetry. In the presence of time-reversal symmetry in the ground state of the superconductor we find the following further constraints on both quantities:
\begin{equation}\label{TRSJandBCD}
    \mathcal{ J}_{\mathbf{q}}=-\mathcal{ J}_{-\mathbf{q}},\quad\mathcal{D}_{\mathbf{q}} = \mathcal{D}_{-\mathbf{q}}.
\end{equation}
As a result, $\mathcal{ J}_{\mathbf{q}}$ vanishes for $\mathbf{q}=0$ but $\mathcal{D}_{\mathbf{q}}$ survives even at $\mathbf{q}=0$ in time-reversal symmetric systems. Note that due its pseudo-vector nature, $\mathcal{D}_{\mathbf{q}}$ in a two-dimensional system requires the breaking of all rotational groups. 

In non-superconducting electronic materials, DC photocurrents act as an e.m.f. that sustain photovoltages across a device enabling to directly measure the photocurrent produced. In contrast, such direct measurements of DC ${j}^\gamma_{\rm photo} [\mathbf{q}]$ in a superconductor are not readily available. To see why, consider the following set-up where a superconducting sample at equilibrium is connected in parallel with an ammeter; the external circuit has a resistance $R_{\rm ext}$ while the superconductor has no resistance. Here without loss of generality, we will focus on the case where the dark equilibrium current is zero. If a uniform DC dissipationless photocurrent generated via Eq.~(\ref{jphoto}) flows into the external circuit $I_{\rm photo}$, it would generate a potential drop across the superconductor $I_{\rm photo} R_{\rm ext}$. However, static electric fields are readily screened in superconductors: superconductors cannot sustain a DC voltage drop. Instead, spatial gradients of superconducting phase (i.e. finite $\hbar q$) will develop to completely screen any voltage drop in the superconductor thereby zeroing the total DC current in the circuit and rendering traditional DC photovoltage/photocurrent measurements inaccessible.

How does $\mathbf{j}_{\rm photo}$ manifest in the response of a superconducting device? As we now argue, $\mathbf{j}_{\rm photo}$ manifests in its effect on the superconductor's current-phase relation. To see this, we average Eq.~(\ref{fullJ}) over one cycle to find the total DC current as 
\begin{equation}
\label{eq:jself}
    \mathbf{j}_{\rm total} (\mathbf{p}) = \langle \mathbf{j} (\mathbf{p},t )\rangle_{\rm cycle} = \mathbf{j}_{\rm dark} (\mathbf{p}) + \mathbf{j}_{\rm photo} (\mathbf{p})
\end{equation}
where $\mathbf{p}$ in the above equation should be understood as the self-consistent center of mass momentum of the superconductor in the presence of irradiation. Note that DC $\mathbf{j}_{\rm total} (\mathbf{p})$ is fixed externally by the total current flowing through the external circuit; as explained above, this current does not change when light is incident on the superconductor. Instead, the center of mass momentum shifts. For small $\mathbf{j}_{\rm photo}$, the self-consistent center of mass momentum can be approximated as 
\begin{equation}
    \mathbf{p} \approx \mathbf{q}_{\rm d} + \delta \mathbf{q}_{\rm photo}, \quad  \delta \mathbf{q}_{\rm photo} = [\boldsymbol{\rho}_{\rm dark} (\mathbf{q}_{\rm d}) ]^{-1} \mathbf{j}_{\rm photo} 
    \label{eq:totalCPR}
\end{equation}
where $\delta \mathbf{q}_{\rm photo}$ is a light-induced superconducting phase gradient and $\mathbf{q}_{\rm d}$ is the center of mass momentum in the dark.

Eq.~(\ref{eq:jself}) readily produces changes to the CPR, see Fig.~(\ref{SDEFIG1}). For example, circularly polarized light induced photocurrent shifts the current phase relationship in a way that depends on the helicity of light, Fig.~(\ref{SDEFIG1})b. Such shifts can have striking effects on a range of observables. A natural example is the critical current. In the dark, supercurrent flows between $[\mathbf{j}_c^-]^{\rm dark}$ and $[\mathbf{j}_c^+]^{\rm dark}$, see blue region in Fig.~(\ref{SDEFIG1})a. Incident light irradiation, on the other hand, alters critical current, see green and red regions in Fig.~(\ref{SDEFIG1})b. Noting that $\mathcal{D}_{\mathbf{p}}$ is even in $\mathbf{p}$ [Eq.~(\ref{TRSJandBCD})] while $\mathbf{j}_{\rm dark}$ is odd [Eq.~(\ref{jdark})],  we find that irradiating with circularly polarized light imbalances the current phase relation $\mathbf{j}_{\rm total}(\mathbf{p}) \neq - \mathbf{j}_{\rm total}(-\mathbf{p})$. As a result, circularly polarized irradiation induces a superconducting diode effect $[\mathbf{j}_c^+]^{\rm total}_{\rm cir} \neq - [\mathbf{j}_c^-]^{\rm total}_{\rm cir}$. Note that the precise value of $\mathbf{j}_c^{+,-}$ is sensitive to the device geometry and can be determined directly from Eq.~(\ref{eq:totalCPR}) numerically, see below for an extended discussion. 

Note that for 2D superconductors, when $C_{3z}$ symmetry is present, $\mathcal{D}_{\mathbf{q}=0}=0$ even when inversion symmetry is broken: finite $\mathcal{D}_{\mathbf{q}=0}$ requires broken rotational symmetry (see below). In this regime we find a counterintuitive observation: ${j}_{\rm photo} [\mathbf{p}\sim 0]\approx \xi \mathcal{D}_{\mathbf{0}}\neq 0$, thus the superconducting system can be tuned into regime where Cooper pairs center of mass momentum $\mathbf{p}$ is {\it opposite} to the supercurrent direction $\mathbf{j}[\mathbf{p}]$. Such a regime of operation indicates the photo-induced component of current dominates over the ordinary supercurrent. 

Lastly we note that both $\mathcal{J}_\mathbf{q}$ and $\mathbf{j}_{\rm dark}$ are odd in $\mathbf{q}$. 
As a result, while linearly polarized irradiation modifies the magnitude of the critical current 
$[\mathbf{j}_c^+]^{\rm total}_{\rm linear} \neq [\mathbf{j}_c^+]^{\rm dark}$, it remains reciprocal $ [\mathbf{j}_c^+]^{\rm total}_{\rm linear} = - [\mathbf{j}_c^-]^{\rm total}_{\rm linear}$.

\vspace{2mm}
{\color{blue}\it Driven superfluid weight.} Beyond changes to the DC critical current characteristics, photocurrent in Eq.~(\ref{eq:totalCPR}) also changes the superconductor's dynamical properties. To appreciate this, we consider a slow probe electric field $\mathbf{E}_{\rm probe} (t) = \boldsymbol{\mathcal{E}}_{\rm probe} e^{-i\omega t} + c.c.$ that probes the driven superconducting state. For slow $ 0 < \omega \ll \omega_d $, the dynamical current response can be described by an imaginary (inductive) linear dynamical conductivity
\begin{equation}\label{linCOND}
    {\rm Im}~\sigma^{\alpha\beta}(\omega) = -\frac{e^2}{\hbar} \frac{\rho^{\alpha\beta}_{\rm tot} (\mathbf{p})} {\hbar\omega},
\end{equation}
where $\rho^{\alpha\beta}_{\rm tot} = (\hbar/e \tau_d) \int _0^{\tau_d} (\partial j^\alpha(\mathbf{q},t)/\partial {q}^\beta) dt $ is the stroboscopic superfluid weight (averaged across one cycle of the drive, $\tau_d = 2\pi/\omega_d$).

\begin{figure*}[t]
    \centering
    \includegraphics[width=0.99\textwidth]{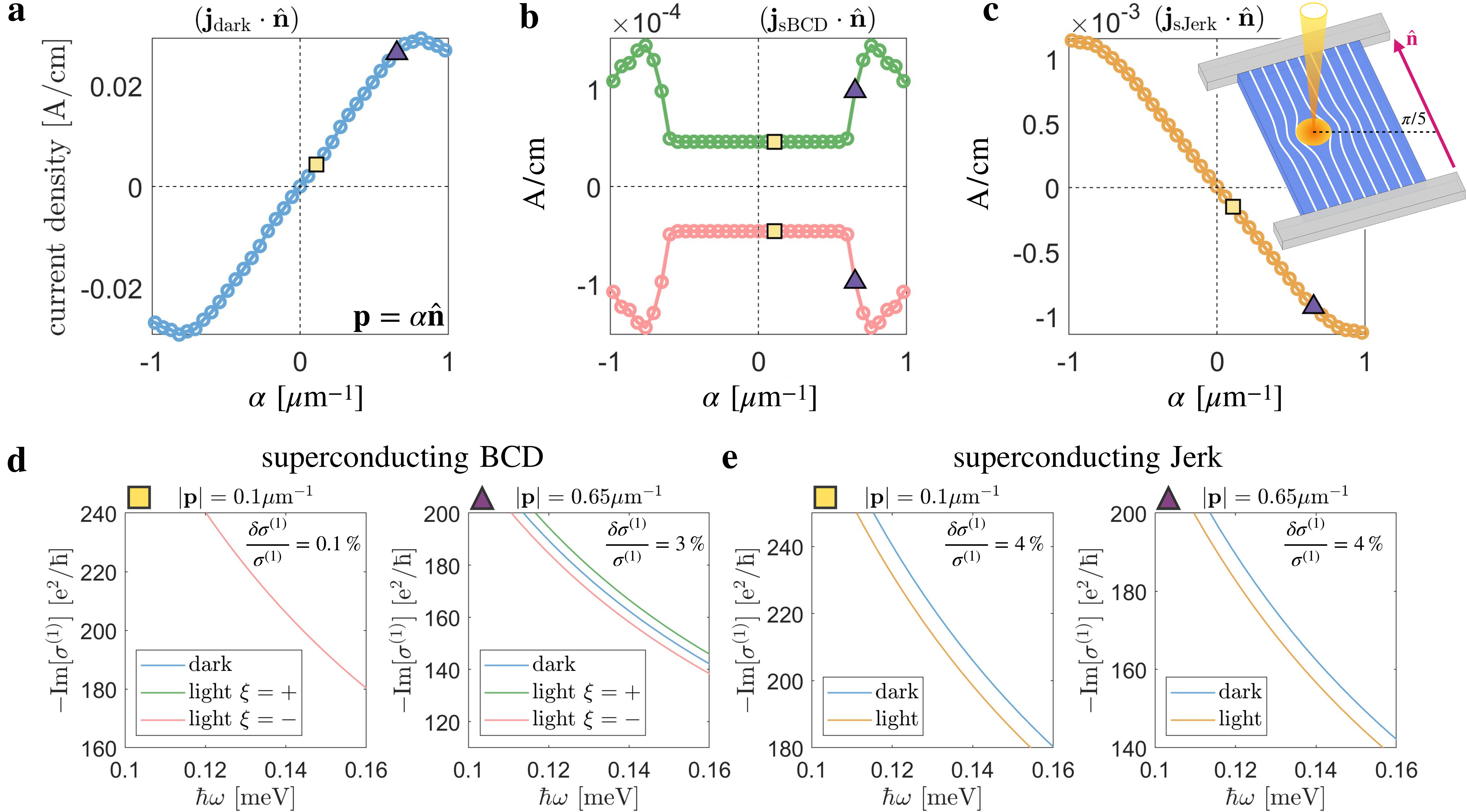}
    \caption{\textbf{Light-enriched superconducting current–phase relationship in dual-gated rhombohedral trilayer graphene (RTG).}   \textbf{(a)} The dark supercurrent [Eq.~(\ref{jdark})] in the absence of light for RTG is odd in center of mass momentum.  \textbf{(b)} The sBCD component of the \textit{light}-induced supercurrent under circularly polarized light, $\boldsymbol{\mathcal{E}} = |\boldsymbol{\mathcal{E}}|(1, \pm i)/\sqrt{2}$ (red for `$+$`, green for `$-$`) and  \textbf{(c)} the sJerk component of the \textit{light}-induced supercurrent, remains the same for both chiralities; both sBCD and sJerk contribution computed using Eq.~(\ref{jphoto}).  All currents in {\bf a-c} are numerically evaluated along a direction forming an angle of $\pi/5$ with the x-axis, where the x-axis is aligned with the zigzag direction of RTG. 
    \textbf{(d)} Contribution of sBCD to the dynamical inductance [imaginary part of the dynamical conductivity] in Eq.(\ref{linCOND}), is sensitive to the light chirality  (switches the sign);  \textbf{(e)} contribution of sJerk to the dynamical inductance is insensitive to the light chirality. See {\bf SI} for parameter values for dual-gated RTG, and we have used light irradiation power of $1\, {\rm W}{\rm cm}^{-2}$ as an illustration. 
 }
    \label{Figcurrentandinductance}
\end{figure*}

Employing Eq.~(\ref{jphoto}), we write the total superfluid weight:
\begin{equation}
\label{eq:lightsuperfluid}
    \rho_{\rm tot}^{\gamma\delta}(\mathbf{p})  = \rho^{\gamma\delta}_{\rm dark}[\mathbf{p}] + \chi^{\gamma\delta\alpha\beta}(\mathbf{p}) \mathcal{E}^{*\alpha} \mathcal{E}^\beta, 
\end{equation}
where the light induced change in the superfluid weight is captured via 
\begin{equation}
   \chi^{\gamma\delta\alpha\beta} _{\mathbf{p}}\hspace{-0.75mm}= \hspace{-0.75mm}\frac{2e^2}{\hbar\omega_d}\bigg\{
   \frac{1}{\hbar\omega_{d}}
   \frac{\partial \mathcal{J}^{\alpha\beta\gamma}_{\mathbf{p}}}{\partial p^\delta}-\frac{i}{2}\frac{\partial\mathcal{D}^{[\alpha\beta]\gamma}_{\mathbf{p}}}{\partial p^\delta} \bigg\},
\label{eq:chi}
\end{equation}
where $\mathcal{D}^{[\alpha\beta]\gamma}_{\mathbf{p}}\equiv \mathcal{D}^{\alpha\beta\gamma}_{\mathbf{p}}-\mathcal{D}^{\beta\alpha\gamma}_{\mathbf{p}}$ and the first term corresponds to a ``snap'' (i.e. third derivative of velocity) characteristic of nonlinear Kerr effects \cite{Ma2025} and the second term describes a Berry curvature quadrupole. 

The stroboscopic changes to superfluid weight in Eq.~(\ref{eq:lightsuperfluid}) demonstrate that driving can control the superfluid weight. Such changes to the superfluid weight can be naturally measured through the low frequency inductive response of superconductors, e.g., by measuring drive induced changes to LC resonance frequencies. This enables to directly track the changes from dark superfluid weight (Fig.~\ref{SDEFIG1}b) to light superfluid weight (Fig.~(\ref{SDEFIG1}d)). 

The transformation properties of $\mathcal{J}_{\mathbf{p}}$ and $\mathcal{D}_{\mathbf{p}}$ play a critical role in Eq.~(\ref{eq:chi}). For example, employing Eq.~(\ref{TRSJandBCD}) for TRS groundstate superconductor, we find that the first term of Eq.~(\ref{eq:chi}) controls non-chiral light-induced superfluid weight, while the second controls circularly polarized light changes to the superfluid weight so that light induced change $\delta \rho^{\gamma\delta}_\mathbf{p} = \rho_{\rm tot}^{\gamma\delta}(\mathbf{p})  - \rho^{\gamma\delta}_{\rm dark}(\mathbf{p})$ as
\begin{equation}
\hspace{-2mm}\delta \rho^{\gamma\delta}_\mathbf{p} \hspace{-1mm}= \hspace{-1mm}{\rm Re} \big[\chi^{\gamma\delta\alpha\beta}_\mathbf{p} \big] {\rm Re} [\mathcal{E}^{\alpha*}\mathcal{E}^\beta] \hspace{-0.4mm}+ \hspace{-0.4mm}{\rm Im}\big[\chi^{\gamma\delta\alpha\beta}_\mathbf{p}\big] {\rm Im} [\mathcal{E}^{\alpha*}\mathcal{E}^\beta],
\end{equation}
where ${\rm sgn} \{{\rm Im} [\mathcal{E}^{\alpha*}\mathcal{E}^\beta]\} = \pm 1, 0$ so that opposite chiralities of light enable to directly switch the superfluid weight in a non-reciprocal fashion: $\rho^{\gamma\delta}_{\rm tot} (\mathbf{p}) \neq \rho^{\gamma\delta}_{\rm tot} (- \mathbf{p})$.  Note that the value of $\mathbf{p}$ can be tuned by running a supercurrent through the circuit: in this way the entire current-phase relationship across different center of mass momenta $\mathbf{p}$ can be mapped and its dependence on quantum geometrical quantities (e.g., Berry curvarture quadrupole) extracted. 

\vspace{2mm}
\textit{\color{blue}{Superconducting photocurrent and light enriched CPR in RTG superconducting heterostructures}.} We now demonstrate how $\mathbf{j}_{\rm photo}$ manifests in a light-enriched superconducting CPR in a two-dimensional noncentrosymmetric superconducting device system. As a simple illustration we consider gated rhombohedral trilayer graphene (RTG) where inversion symmetry breaking can be controlled by dual-gate. Superconductivity can be found in its intrinsic state \cite{Zhou2021} or proximitized. Here we take on an effective approach \cite{matsyshynSBCD,LevitovSC,Zhang_2010,Jung_2013} and focus on a simple mean-field description with broken inversion symmetry as well as broken $C_{3z}$ symmetry, see {\bf Supplementary Information} for description. 

We plot the DC dark current [Eq.~(\ref{jdark})] and superconducting photocurrent [Eq.~(\ref{jphoto})] in Fig.~\ref{Figcurrentandinductance}a-c as a function of center of mass momentum. Here red Fig.~\ref{Figcurrentandinductance}b describes the photocurrent from $\mathcal{D}^{\alpha \beta \gamma}_{\mathbf{p}}$ (superconducting Berry curvature dipole) and green Fig.~\ref{Figcurrentandinductance}c describes photocurrent from $\mathcal{J}^{\alpha \beta \gamma}_{\mathbf{p}}$ reflecting their even and odd character under $\mathbf{p} \to - \mathbf{p}$; indeed, Berry curvature dipole is nonzero even when $\mathbf{p} = 0$ (reflecting the broken $C_{3z}$) while the superconducting Jerk contribution requires a finite $\mathbf{p} \neq 0$ to activate. As explained above, such superconducting photocurrent cannot be directly detected. Instead, they manifest in how the superconducting CPR is altered. For e.g., in the presence of circularly polarized induced superconducting photocurrent (red), we find that the inductance Eq.~(\ref{linCOND}) becomes non-reciprocal as plotted in Fig.~\ref{Figcurrentandinductance}d and can be switched for different polarizations of light $\xi =\pm 1$. These are associated with ${\rm Im}[\chi]$ and the superconducting Berry curvature quadrupole. Interestingly, we find appreciable changes of several percent can be achieved even for modest irradiation of 1$\rm W/cm^2$. While the changes are smaller for small $\mathbf{p}$ values [see e.g, yellow square], these changes are most pronounced at large $\mathbf{p}$ values [black triangle] close to the critical momentum. Indeed, this reflects the small Berry curvature quadrupole values at small $\mathbf{p}$ but appreciable values at larger $\mathbf{p}$ demonstrating how such a light-induced inductance can enable to map out the quantum geometric properties of the superconductor. 
 
In contrast, for linearly polarized light, the changes to the superfluid weight are reciprocal [(see ${\rm Re}[\chi]$ in Eq.~(\ref{eq:chi})] and appreciable even at small $p$ values: they come up to several percent even for modest values of incident light. Notice that the light-induced inductance for linearly polarized light remain is relatively insensitive to $\mathbf{p}$ reflecting the near constant slope of the superconducting Jerk photocurrent. 

We have demonstrated how superconducting photocurrent drives changes to its current phase relation. This manifests in light-driven changes to a range of fundamental superconducting devices properties that include its inductance as well as its critical current. While in time-reversal invariant noncentrosymmetric superconductors, finite $\mathcal{D}^{\alpha \beta \gamma}_{\mathbf{p}=0}$ (and its associated circular superconducting photocurrent) requires a broken rotational symmetry, we find $\mathcal{D}^{\alpha \beta \gamma}_{\mathbf{p}\neq0} \neq 0$ even for $C_{3z}$ symmetric materials. This means that light induced non-reciprocal inductance we discuss can persist at finite $\mathbf{p}=0$. This is particular important for the critical current where circularly polarized superconducting photocurrent can induce a diode effect that persists even in a $C_{3z}$ preserving superconducting device. Non-dissipative superconducting photocurrent provides light-induced means of controlling superconducting device properties as well as protocols for measuring superconducting photocurrent in devices. Since superconducting photocurrents can be high sensitive to the symmetry of the order parameter, we expect that this protocol can be used not only as a probe of the Cooper pair momentum dependendent quantum geometry of the superconducting state \cite{matsyshynSBCD}, but also the structure of the its order parameter. 
 
\textit{\color{blue}Acknowledgements.} We gratefully acknowledge useful conversations with Valla Fatemi, Giovanni Vignale, Roshan Krishna Kumar. This work was supported by Singapore Ministry of Education Tier 2 grant MOE-T2EP50222-0011 (JCWS)

\bibliography{CPR}
\clearpage

\appendix
\setcounter{equation}{0}
\renewcommand{\theequation}{S\arabic{equation}}
\renewcommand{\thefigure}{S\arabic{figure}}
\renewcommand{\thetable}{S\Roman{table}}
\titleformat{\section}[block]{\normalfont\normalsize\bfseries\centering}{Section \Alph{section}:}{0.3em}{}

\onecolumngrid
\section*{\large Supplementary Information for ``Superconducting Photocurrent and Light Enriched Supercurrent Phase Relation''}
\subsection*{Superconducting RTG model details}
In this supplement we detail the low-energy effective model used for computing the photocurrent and superconducting current phase relation (CPR) in both the dark and light regimes for dual-gated rhombohedral trilayer graphene (RTG) in the main text. In modeling RTG we employed an effective low-energy effective two-band Hamiltonian; this can be obtained via a decomposition to $(\ket{A_1},\ket{B_3})$ basis from the full six-band hamiltonian of RTG, see e.g., Ref.\cite{Zhang_2010,Jung_2013,matsyshynSBCD}. This two band effective model reads: 
\begin{multline}\label{ABC}
    \hat{H}_{\rm RTG}^{(0)}(\mathbf{k},\xi)=\frac{v_0^3}{\gamma_1^2}\left(\begin{array}{cc}
        0 &(\pi_\xi^\dagger)^3  \\
        (\pi_\xi)^3 & 0
    \end{array}\right)+\frac{v_0\gamma_N }{\gamma_1} \left(\begin{array}{cc}
        0 &\pi_\xi^\dagger  \\
         \pi_\xi & 0
    \end{array}\right)+\delta_2\left(1-3\frac{k^2v^2_0}{\gamma^2_1}\right)\tau_0+\delta\left(1-\frac{k^2 v_0^2}{\gamma_1^2}\right)\tau_z\\+\left(\delta_1-\frac{2v_4v_0k^2}{\gamma_1}\right)\tau_0 +\left(\frac{1}{2}\gamma_2 -\frac{2k^2v_0v_3}{\gamma_1} \right)\tau_x, 
\end{multline}
with $\pi_\xi = \xi k_x + ik_y$, $k=|\mathbf{k}|$, $\xi = \pm1$ is the valley index and the $\gamma_N(\mathbf{k})=0.02e^{- v_0 k/\gamma_1}{\rm ~eV}$ is strain. Parameters are adopted from Refs.\cite{Zhang_2010,Jung_2013} as: $\gamma_0=3.1 {\rm~ eV},\gamma_1=0.38 {\rm~ eV},\gamma_2=-0.015, \gamma_3=0.29 {\rm~ eV},\gamma_4=0.141 {\rm~ eV};\gamma_1/v_0 = 0.0573\text{\AA}^{-1}$, $v_i = {\sqrt{3} a \gamma_i}/{2},~a = 2.46 \text{\AA},\delta = 0.03{\rm ~eV},\delta_1= -0.0105{\rm ~eV},\delta_2= -0.0023{\rm ~eV}$. Note, here we suppressed $\hbar$'s for shortness (to restore $\hbar$, replace $v_i\rightarrow v_i/\hbar$). This Hamiltonian preserves the time-reversal symmetry, while $\delta$ can be controlled by external gate potential breaks inversion symmetry and the strain parameter  $\gamma_N$ breaks $C_{3z}$ \cite{Jung_2013}. 

In modeling the superconductivity in RTG, we adopted the standard mean field Bogoliubov-de-Gennes (BdG) description of a superconductor with a local attractive interactions $\hat{V}_{\rm el-el}^{ss'}(\mathbf{r}-\mathbf{r}')=-U \delta(\mathbf{r}-\mathbf{r}')\delta_{s,\bar{s}'}$, where $s,\bar{s}$ are the opposite spins. Finite center of mass momentum (e.g., induced by running a supercurrent through a superconducting device) via an order parameter with a real space dependence as $\Delta(\mathbf{r})=\Delta e^{2i\mathbf{p}\cdot\mathbf{r}}$. We employed a local pairing~Ref.~\cite{LevitovSC} between the electrons of opposite momenta ($\mathbf{k}$ of $K_+$ valley with $-\mathbf{k}$ of $K_-$ valley). Note our treatment is agnostic to the mechanism for attraction. The effective Bogoliubov-de-Gennes (BdG) Hamiltonian for each valley with self-consistently adjusted Cooper pair center of mass momentum $2\hbar\mathbf{p}$ reads as:
\begin{equation}\label{ABCBdG}
    \hat{H}^{\rm BdG}_{\mathbf{k,q},0}(\xi) = \left( \begin{array}{cc}
        \hat{H}_{\rm RTG}^{(0)}(\mathbf{k+p},\xi)-\mu\tau_0 & \hat{\Delta} \\
        \hat{\Delta}^\dagger & -\hat{H}_{\rm RTG}^{*(0)}(\mathbf{-k+p},-\xi) +\mu\tau_0
    \end{array}\right).
\end{equation}

We examined superconducting RTG for $\mu\approx 17$ meV (point of high DoS) and a cut-off frequency of $\hbar\omega_D = 40$ meV. For illustration we used $U = 60$meV yielding a modest $T_c\approx 1$ K. The order parameter is determined self-consistently in a Hartree-Fock fashion by solving $\Delta_{\mathbf{q},ab} = -U\sum_{\mathbf{k}}\langle c_{-\mathbf{k}+\mathbf{q},a}c_{\mathbf{k}+\mathbf{q},b}\rangle$ with $a,b$ to be the lattice degrees of freedom. Self consistently determined order parameter at $T =0, \mathbf{p}=0$ K is $\Delta_A \approx 1.5\cdot10^{-1}$ meV, $\Delta_B \approx 10^{-2}$meV. We find the gap opening in the spectrum of the BdG Hamiltonian of order $\Delta_{\rm SC} \approx 2\cdot10^{-1}$ meV. 

Using the self-consistent order parameter described above, we directly solve Eq.~(\ref{ABCBdG}) for its Bogoloiubov eigenstates, denoted as $\ket{{\Psi}^a_{\mathbf{k,q}}}$, and eigenenergies, $\varepsilon_{\mathbf{k},a}(\mathbf{q})$. To compute the \textit{light}-induced light-induced photocurrents in $\mathbf{j}_{\rm photo}$ we use the definitions of the superconducting Berry curvature dipole and superconducting Jerk terms provided in the main text after Eq.~(\ref{jphoto}). In order to plot the current in Fig.2 of the main text, we chose a coordinate system wherein the x-axis is defined along the zigzag direction. For the current, we employed a fixed orientation of the Cooper pair momentum vector, $\angle(\mathbf{p}, x) = \angle(\mathbf{n}, x) = \pi/5$.
\end{document}